# H$_2$ Mapping on Pt-loaded TiO$_2$ Nanotube Gradient Arrays

G. Loget,[†] P. Schmuki[*,†,‡]

[†]Department of Materials Science WW-4, LKO, University of Erlangen-Nuremberg, Martensstrasse 7, 91058 Erlangen, Germany

[‡]Department of Chemistry, King Abdulaziz University, Jeddah, Saudi Arabia.

* Corresponding Author. E-mail: schmuki@ww.uni-erlangen.de

**ABSTRACT**: We describe a rapid screening technique for determining the optimal characteristics of nano- photocatalysts for the production of H2 on a single surface. Arrays of TiO2 nanotubes (NTs) with a gradient in length and diameter were fabricated by bipolar anodization and a perpendicular gradient of Pt nanoparticles (NPs) was generated by the toposelective decoration of the TiO2 NTs. Photocatalytic hydrogen evolution was locally triggered with a UV laser beam and the arrays were screened in x- and y-direction for spatially resolved kinetic measurements and the mapping of the optimal hydrogen production. By using this technique, we demonstrate the time-efficient and straightforward determination of the tube dimensions and the Pt amount for an optimized H2 production. The concept holds promise to generally improve the study of many photoreactions as a function of the physicochemical characteristic of nano-photocatalysts, which renders it highly attractive for the optimization of various important chemical processes.

## Introduction

Over the last decades, tremendous efforts have been made to improve the conversion of light into fuel by photocatalysis or photoelectrochemistry.1-7 A common path-way to H2 fuel typically requires a semiconductor surface that, upon light absorption, generates electron-hole pairs which react with a source for protons. Over the past few years, arrays of self-assembled $TiO_2$ nanotubes (NTs) in various modifications (doped, decorated and junctions) are one of the most investigated semiconductor surfaces for photocatalysis because they combine the intrinsic benefi-cial electronic properties of $TiO_2$ with highly controllable geometry.8,9

The H2 evolution on surfaces of $TiO_2$ NT arrays can be accelerated by sacrificial co-solvents as hole scavengers (e.g. methanol or ethanol) and by the decoration of the surface with co-catalysts such as noble metal nanoparticles (NPs).10-28 In particular, it is well established that Pt NPs, attached to the $TiO_2$ nanostructure, drastically enhance the H2 production due to electronic and catalytic effects, and thus allow obtaining reasonable amounts of H2 under un-biased conditions.27,28 In the use of nanotubular $TiO_2$, H2 production also strongly depends on the $TiO_2$ NTs dimen-sions (namely length and diameter), since they control crucial parameters such as the surface area, the diffusion of ions and electrons inside the tubes, as well as light ab-sorption and scattering.8,9

Although different strategies have been explored for decorating $TiO_2$ NTs with Pt NPs,25-30 only little focus has been given so far to the systematic investigation of the tube dimensions on the hydrogen evolution yields. There-fore, in this work we introduce a reliable, time-effective and versatile analytical method that allows studying the H2 production as a function of the tube dimensions and the amount of co-catalyst immobilized on the tubes.

The fabrication of self-assembled $TiO_2$ NTs arrays is typically achieved by anodization of a Ti foil in a viscous, fluoride containing electrolyte. This leads to homogeneous $TiO_2$ layers with tubes of the same length and diameter (Figure 1a).8,9 A broad screening of tube dimensions with this technique is time-consuming, because it requires fabricating and studying series of individual surfaces. A more efficient approach to fabricate graded $TiO_2$ NTs arrays is based on bipolar electrochemistry, a wireless phenomenon that allows triggering of redox reactions on the opposing ends of a conducting object.33,34 Bipolar electrochemistry takes place when a sufficiently high electric field polarizes a conducting object and leads to the flow of a fraction of the delivered current within the object.33,34 This faradaic current induces spatially-separated reduction and oxidation reactions along the surface of the conducting object that are more pronounced at the extremities, where the polarization is maximum. This phenomenon is currently attracting a considerable attention in the domains of analytical chemistry,35-43 materials science,44-47 particle mo-tion48,49 and seawater desalination.50 We recently reported the controlled fabrication of $TiO_2$ NT gradients with de-fined length and diameter by bipolar anodization (Figure 1b).51

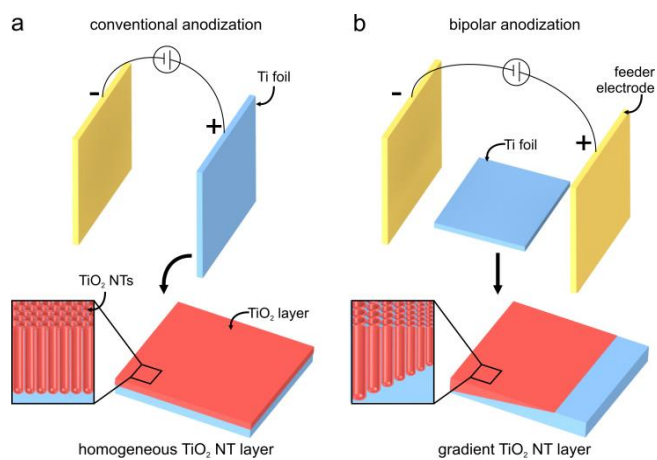

Figure 1. Fabrication of TiO2 NT gradients by bipolar anodization. a) Conventional anodization leads to homogenous TiO2 NT layers with the same length and diameter. b) Bipolar anodization leads to TiO2 NT layers with a gradient in tube length and diameter.

Here we demonstrate the fabrication of opened and catalytically active TiO2 NT gradients by a combination of bipolar anodization and Pt decoration. We then show the spatial mapping of hydrogen evolution on these NT arrays. This approach allows to determine on a single surface the optimal tube geometry and Pt amount for an optimized H2 photoproduction.

Experimental

Chemicals and materials

Chemicals and titanium foils (0.125 mm thick, >99.6 % purity, Advent, England) were used as received. Ethylene glycol (>99.5 %) was purchased from Fluka, ethanol (99.5 %), acetone (Reag. Ph. Eur.) and methanol (Reag. Ph. Eur.) were purchased form Merck. Ammonium fluoride (ACS reag. >98 %) was purchased from Sigma Aldrich. The electrolyte was prepared with ultrapure water (resistivity = 17.1 mΩ.cm).

Fabrication of the TiO2 NT gradients

The electrolyte was composed of ammonium fluoride (0.1 M) and water (5 wt%) in ethylene glycol. Titanium pieces of 1.5 x 4.2 cm were cut, degreased by sonication in acetone, ethanol and water, and dried under a nitrogen stream. The fabrication procedure was performed as reported recently.51 The electrochemical cell was composed of a plastic beaker in which a Teflon support has been fixed in the bottom. In order to avoid any motion of the sample during the electrolysis as well as electrochemistry on the bottom of the foil, the Ti surface was tapped on a Teflon detachable support (that was designed to fit on the fixed support) with Kapton tape covering 1 mm of its edges. The detachable support was placed on the fixed support. The electrolyte ≈140 mL was then poured in the cell and the two feeder electrodes (platinum foils, 1.2 x 2.1 cm) were placed at a distance of 1 mm from the uncovered Ti foil edges. The electrolysis was performed under stirring. The electric field of 21.4 V.cm-1 (E = U/d, with E being the electric field, the imposed potential U = 90 V and the distance between the feeder electrodes d = 4.2 cm) was applied for 50 min using a LAB/SM 1300 power source (ET System). After the electrolysis, the movable support was placed in ethanol overnight in order to remove the fluorides from the nanotubes and to detach the Kapton tape from the surface. The titanium foil was then rinsed with ethanol and water, and dried under a nitrogen stream. The samples were annealed using a rapid thermal annealer (Jipelec Jet-First100) at 450 °C for 1 h in air, before and after this constant temperature step, the heating and cooling rates were 30°C min-1. For removing the nanograss from the tube tops, the samples were sonicated in water using a Elmasonic P (Elma) sonicator for 5 minutes and were rinsed under running tap water, this was repeated tree times, and then the samples were rinsed with ultrapure water and dried under a nitrogen stream. The decoration with Pt NPs was done by sputtering Pt with a high vacuum sputter coater EM SCD 500 (Leica) using a sputtering rate of 0.08 nm.s-1. First, 1 nm of Pt was sputtered on the whole gradient (as defined by the quartz crystal of the sputter coater) by sputtering for 12 s and then a Ti sheet was used for covering a 0.5 x 4 cm area of the surface. 9 nm of Pt were further sputtered during 113 s and the Ti mask was placed over a 1 x 4 cm area (comprising the first protected area). Finally, 10 nm of Pt were sputtered during 125 s. This masking/sputtering procedure allowed to deposit 1, 10 and 20 nm (values of the compact Pt layer deposited on the quartz crystal) on 3 separated areas of 0.5 x 4 cm. X-ray diffraction analysis (XRD, Xpert Philips PMD with a Panalytical X'celerator detector) using graphite monochromized CuKα radiation was used for determining the crystal structure of the samples. The X-ray photoelectron spectroscopy (XPS) was performed with a PHI 560 XPS system. All the SEM characterizations were performed using a FE 4800 SEM (Hitachi).

Hydrogen production and analysis

For the hydrogen production experiments the gradient surface was placed in a cylindrical quartz cell that was filled with 10 mL of the water/methanol mixture (80/20 vol%). The cell was placed on a micropositioner stage (Karl Süss) mounted on a stand (see Figure 5). The cell was purged by bubbling N2 gas for 20 minutes and sealed with septums. The laser beam (Kimmon IK Series, 200 mW) was directed on the surface for a desired time before shut-ting it off. After the experiment, a sample of 0.2 mL of the dead volume was taken out with a syringe and was ana-lyzed by a GC-2010 plus gas chromatograph (Shimadzu). For the kinetic measurements, GC measurements were performed every 15 min without replacing and purging the solution in the cell. For the mapping, the production was carried out for 15 min for each point and the solution was replaced and purged after each measurements. In this case, the gas analyses were repeated three times after each run to ensure a good reproducibility. The total time of analysis ta for each Pt amount during the mapping can be calculat-ed by ta = N (tp + till + 3tGC), with N being the number of illuminated spots, tp the purging time, till the illumination time and tGC the gas analysis time. Typically, with N = 5, tp = 20 min, till = 15 min and tGC = 6 min, ta = 265 min.

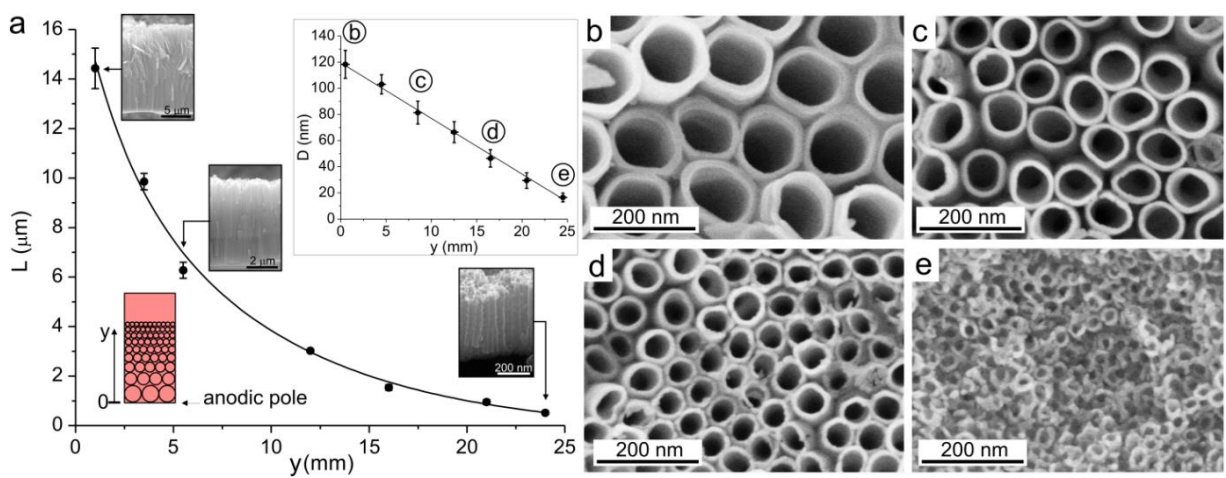

Figure 2. Gradients of opened TiO2 NTs. a) Plot of the tube length L as a function of the y-position of the gradient. The SEM pictures are cross-sections taken at the y-positions indicated by the arrows. The bottom left scheme depicts a top view of a TiO2 NTs gradient and defines the y-axis. Inset, top

right: Plot of the tube inner diameter D as a function of the y-position. b-e) SEM pictures showing top views of the tubes at the y-positions b-e indicated in the inset (top right) of Figure 2a.

Results and Discussion

The TiO2 NT gradients, as in Figure 2, were fabricated by bipolar anodization.51 In our case, a Ti foil (1.5 x 4.2 cm) was placed between two feeder electrodes in a cell with a fluoride-containing electrolyte. An electric field of 21.4 V.cm-1 was applied between the feeder electrodes for 50 min while stirring continuously. Under these conditions the Ti foil was sufficiently polarized to act as a bipolar elec-trode, which means that it behaved at the same time as an anode and as a cathode. The reduction of the electrolyte occurred at the edge of the foil facing the feeder anode (the cathodic pole of the bipolar electrode) together with the oxidation of the Ti on the edge of the foil facing the feeder cathode (the anodic pole of the bipolar electrode). At the anodic pole of the bipolar electrode, two competing oxida-tive processes proceeded: the formation of TiO2 and the dissolution of fluoride-complexed Ti-species. This resulted in the formation of a layer of self-assembled TiO2 NTs with a gradient in length and diameter.8,9,51 Both redox processes can be monitored in-situ at the bipolar electrode, the elec-trolyte reduction at the cathodic pole of the bipolar elec-trode provokes gas evolution and the formation of TiO2 layers at the anodic pole induces a color change.

The so-created layers of TiO2 NTs gradients spread over more than 60% of the surface of the foil. Scanning electron microscope (SEM) observation of the top of these layers revealed that the tubes were covered by a layer of nanograss (see Figure S1a), caused by the etching of the opening of the tubes by the fluorides during the anodiza-tion.8,9 This nanograss is typical for Ti anodization but undesirable since it can impair the decoration of the tube walls or the diffusion within the tubes. It was therefore removed from the tube tops by an ultra-sonication treat-ment (see experimental section for details), which led to smooth NT openings, which are shown in high magnifica-tion in Figures 2b-e and in low magnification in Figure S1b.

The top view and different cross-sections of the resulting TiO2 NTs layers were analyzed by SEM. Figure 2a shows the variation of tube length L and inner diameter D with the length of the foil (along its y-axis). The tubes with the highest L and D values were located at the edge of the anodic pole and both values decreased along the y-axis. The tube length decreased exponentially from Lmax = 14.4 µm to Lmin = 0.6 µm. The tube diameter decreased linearly (see the inset of Figure 2a and in Figures 2b-e) from Dmax = 118 nm to Dmin = 17 nm. These variations in tube length and diameter along the NT layer are in good agreement with our previous report and are explained by the respec-tive current and potential distributions over the anodic pole during the bipolar anodization.51 These results show that layers of TiO2 NT gradients with opened tubes can be fab-ricated by bipolar anodization. In the following, we will discuss how to make these gradients photocatalytically active for the production of H2.

For this, we treated the surfaces according to the proce-dure shown in Figure 3a. First, the tubes were converted into TiO2 anatase by thermal annealing of the gradient at 450 °C in air. The arrays were then decorated with Pt NPs by sputtering. In order to toposelectively decorate the gra-dients, three sputtering steps were successively performed. For the first one, Pt was sputtered for 12 s (sputtering rate = 0.08 nm.s-1) on the whole surface, which corresponded to a Pt layer with a nominal thickness TPt = 1 nm, as measured by the quartz crystal of the sputter coater. Before the sec-ond sputtering step, a mask was placed on an rectangular area of 0.5 x 4 cm for protecting this surface and the sput-tering was performed during 113 s (TPt = 9 nm). Finally, the mask was placed on a 1 x 4 cm area (which comprised the first protected area) and the sputtering was performed for 125 s (TPt = 10 nm). This sputtering/masking procedure allowed decorating three separated areas with different sputtering times ts (12, 125 and 250 s).

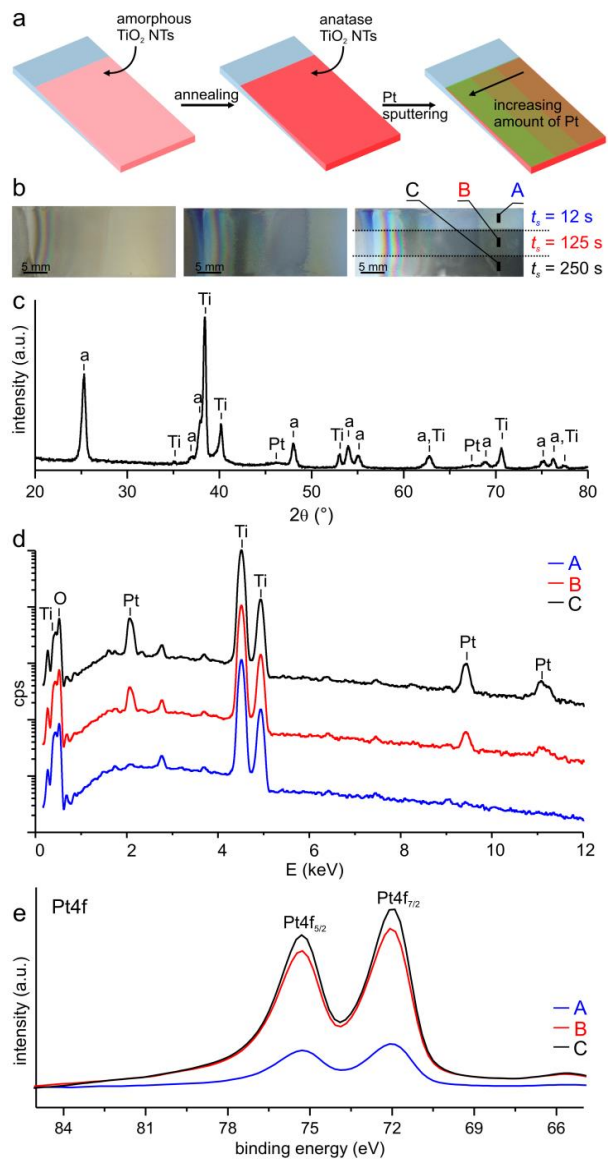

Figure 3. Annealing and decoration of the TiO2 NT gradients with three different amounts of Pt. a) Scheme showing the modification steps performed on the TiO2 gradients. b) Photographs of: an as-formed opened TiO2 NTs gradient (left), an annealed gradient (middle) and an annealed gradient decorated by sputtering Pt during three times (12, 125 and 250 s) (right). c) XRD spectrum taken on a Pt-decorated gradient surface (a = TiO2, anatase). d) EDX spectra taken at the position A, B and C indicated in Figure 3b. e) XPS spectra showing the Pt 4f doublet at the position A, B and C indi-cated in Figure 3b.

Figure 3b shows that the thermal treatment considerably changed the color of the TiO2 NTs gradient, from mostly grey before annealing to mostly bluish after annealing. After sputtering, the three areas decorated with different Pt amounts could be visually distinguished by their lightness: the darkest area had been modified for the longest time and the lightest area for the shortest time (see Figure 3b).

The crystalline structure of the modified surface was an-alyzed by X-ray diffraction (XRD), as shown in Figure 3c. This spectrum reveals the characteristic peaks of TiO2 ana-tase, Ti and two small peaks for Pt ($2\theta$ = 46° and 67°), no rutile peak was observed, which is desirable because ana-tase has superior electronic properties compared to rutile.[8]

The characterization of the amount of Pt in each specific areas was performed by energy dispersive X-ray spectrom-etry (EDX), X-ray photoelectron spectroscopy (XPS) and SEM, at the positions A, B and C indicated in Figure 3b. First, the bulk composition of the layer was characterized by EDX. The three spectra, shown in Figure 3d, revealed the Ti and O peaks - the peak at 2.8 keV being the escape peak of Ti-K$\alpha$. Pt could be detected by EDX at the posi-tions B and C where ts = 125 s and 250 s, respectively. For position A where ts was the smallest (12 s), the detection limit did not allow to observe Pt. Nevertheless, Pt on the surface of the tubes could be clearly observed for the three positions A, B and C by XPS as shown by the Pt 4f dou-blets of Figure 3e. The positions of these peaks correspond to the ones of metal Pt.[52]

Analysis of the EDX and XPS Pt peaks allowed to de-termine semi-quantitatively the Pt amounts relative to the bulk and on the surface of the TiO2 NTs, respectively, for each position. The values, reported in Table 1 shows that the Pt amount determined by EDX varied from 0 to 5.1 wt% and the one determined by XPS varied from 71.6 to 92.8 wt%, from the area having the minimum Pt loading (ts = 12 s) to the area having the maximum Pt loading (ts = 250 s). The values are in good agreement with an increas-ing loading of Pt on the TiO2 NTs for increasing sputtering time.

Morphological investigation of the Pt was performed by SEM observations at the positions A, B and C, that are shown in Figure 4 a-c, respectively. The left row of Figures 4 a-c reveals that the top wall thicknesses of the TiO2 NTs (the values are reported in Table 1) increased with the amount of deposited Pt from 13 ±3 nm for ts = 12 s to 23 ±3 nm for ts = 250 s. SEM pictures of cross-sections for the three positions at a layer depth of ≈400 nm are shown in the right row of the Figures 4 a-c. They reveal that for the tree observed positions A, B and C, the TiO2 NT walls were decorated with Pt NPs, which increased in size and density with an increasing sputtering time. The penetration depth of the Pt NPs inside the layer was found to not change significantly with the sputtering time but to be dependent on the diameter of the tubes. As shown in Figure S2 and table S1, it varied from a few micrometers for the widest tubes to half a micrometer for tubes with D ≈ 50 nm.

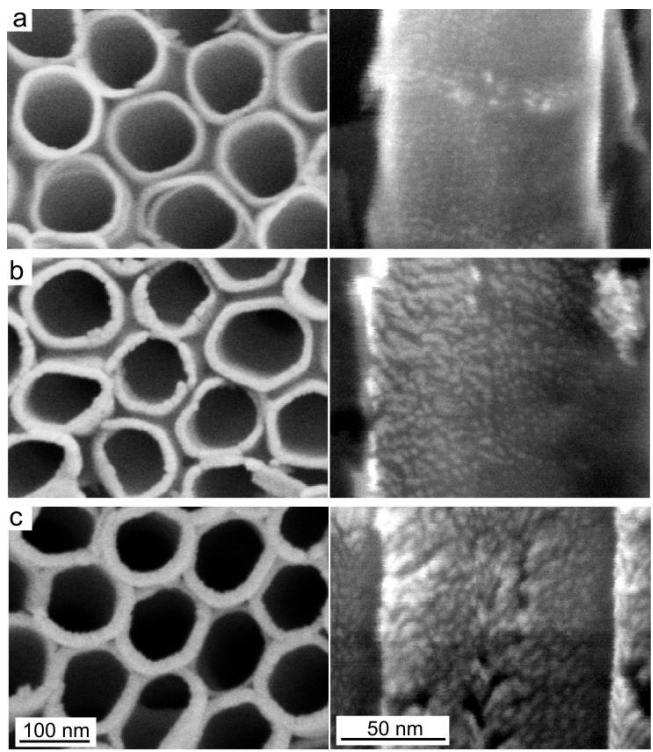

Figure 4. SEM pictures showing top views (left) and tube walls (right) at the positions A (a), B (b) and C (c) indicated in Figure 3b.

Table 1. Pt loading at the positions A, B and C indicated in Figure 3b. Values of sputtering time $t_s$, sputtered thickness $T_{Pt}$, Pt amount determined by EDX and XPS and top wall thickness $T_w$.

|   | $t_s$ (s) | $T_{Pt}$ (nm) | EDX Pt amount (wt%) | XPS Pt amount (wt%) | $T_w$ (nm) |
|---|---|---|---|---|---|
| A | 12 | 1 | 0 | 71.6 | 13 ±3 |
| B | 125 | 10 | 2.1 | 90.5 | 20 ±5 |
| C | 250 | 20 | 5.1 | 92.8 | 23 ±3 |

These characterizations demonstrate that open $TiO_2$ NT gradients can be converted to anatase and toposelectively decorated with Pt NPs. Since these NT surfaces exhibit a gradient in tube length and diameter in y-direction as well as a gradient in Pt NP in x-direction, they appear to be optimal substrates for studying catalytic activities as a function of the individual tube characteristics and the amount of Pt. In the following we will demonstrate the localized production of $H_2$ with these Pt-decorated $TiO_2$ NT gradients.

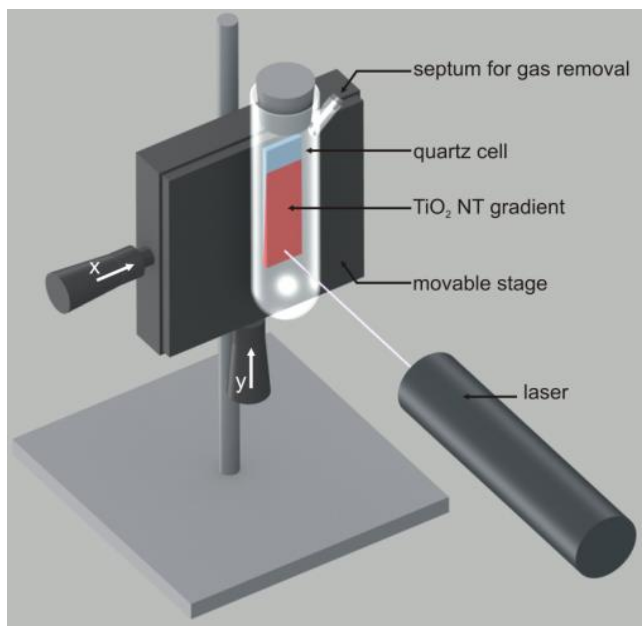

Figure 5. Set-up used for the localized hydrogen production.

Figure 5 illustrates the set-up used for the hydrogen pro-duction experiments. A quartz cuvette containing 10 mL of a water/methanol mixture and the tube gradient was placed on a micropositioner stage in order to scan the gradient along the x-and y-axis. The chosen water/methanol mixture (80/20 vol%) has been reported to optimally accelerate the H2 production.24,27 The cuvette was purged with N2 gas in order to remove the air that was present in the solution and in the dead volume, and was then sealed with septums. A 325 nm laser with a spot size of 1.5 mm (FWHM of the Gaussian intensity distribution) was used to locally illuminate the gradient at a defined x, y-position for a desired time. Then the laser was shut and a sample of the dead volume was analyzed by gas chroma-tography (GC).

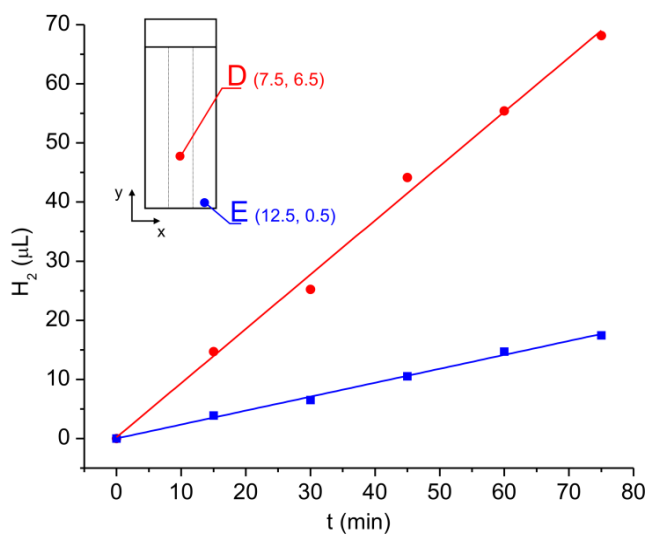

Figure 6. Kinetic measurements of the hydrogen production at 2 positions on a TiO2 NT gradient. Position D : x = 7.5 mm, y = 6.5 mm (red curve) and position E : x = 12.5 mm, y = 0.5 mm (blue curve).

In a preliminary experiment we tested our set-up regard-ing its ability to measure reliably and accurately the H2 production on the NT gradient. Therefore we decided to perform kinetic measurements, shown in Figure 6. Two x, y-positions: D (x = 7.5 mm, y = 6.5 mm) and E (x = 12.5 mm, y = 0.5 mm) localized in areas of different Pt amount (ts = 125 s for D; ts = 12 s for E) were chosen. GC analysis (chromatograms shown in Figure S3) was carried out every 15 min for 75 min in total. The H2 production was found to increase linearly with the illumination time with coefficients of determination r² = 0.9945 and 0.9952 for D and E, re-spectively. As shown by previous reports, linearity is ex-pected for this time range.22,26,27 The two hydrogen produc-tion rates were 55.1 µL.h-1 for position D and 14.3 µL.h-1 for position E. These values are in the same order of mag-nitude as the rates reported for hydrogen production on TiO2 NTs decorated with Pt NPs in water/methanol mix-tures.27,28 These data show that H2 production rates at de-fined positions can be reliably measured with this experi-mental configuration. In the following we will show the mapping of H2 production with this set-up.

For the mapping of hydrogen, the gradient was used in the configuration previously described (Figure 5) and screened along the x- and y-axis. As we have shown be-fore, the hydrogen production increases linearly with time. Therefore the hydrogen production rate could be deter-mined by performing the photocatalysis for 15 min for each spot. Figure 7a shows the H2 mapping that was ob-tained by analyzing 15 spots spaced by 5 mm on the x-axis and 3 mm on the y-axis. This plot illustrates straight-forwardly the spatial resolution of the catalytic activity of the Pt-decorated TiO2 NT gradients for H2 production. First, it is revealed that the H2 production rate follows the same trend at the three areas of different Pt amounts (x-values) along the y-axis with a roughly normal distribution and a maximum value at y = 6.5 mm. This plot reveals that the tubes located at x = 7.5 mm showed the best performance whereas the tubes at x = 2.5 mm and 12.5 mm had weaker H2 production rates, with a minimum for x = 12.5 mm. The maximum hydrogen production rate was found to be 58.9 µL.h-1 and the minimum was 17.4 µL.h-1.

This spatial information can be translated into physico-chemical information since every x, y couple is assigned to a distinct combination of Pt amount and NT dimensions. Like this the Pt amount can be ranked for every given NT dimension regarding its efficiency in H2 production, which is maximal for a Pt sputtering time of 125 s, followed by 250 s and then 12 s. The y-axis screening allows screening the H2 production as a function of the dimension of TiO2 NTs for a constant amount of Pt. The values of L and D along the gradient, previously determined in Figure 2a, allowed us to plot the H2 production rate as a function of L (Figure 7b) and D (Figure 7c). These two graphs make a straightforward determination of the best tube dimensions possible for H2 production, i.e. the best performance was obtained for L = 6 µm and D = 92 nm, values that are well in line with previously published works.53 These results demonstrate that this method rapidly allows the determina-tion of the best characteristics of decorated TiO2 NTs for an optimized H2 production on a single surface.

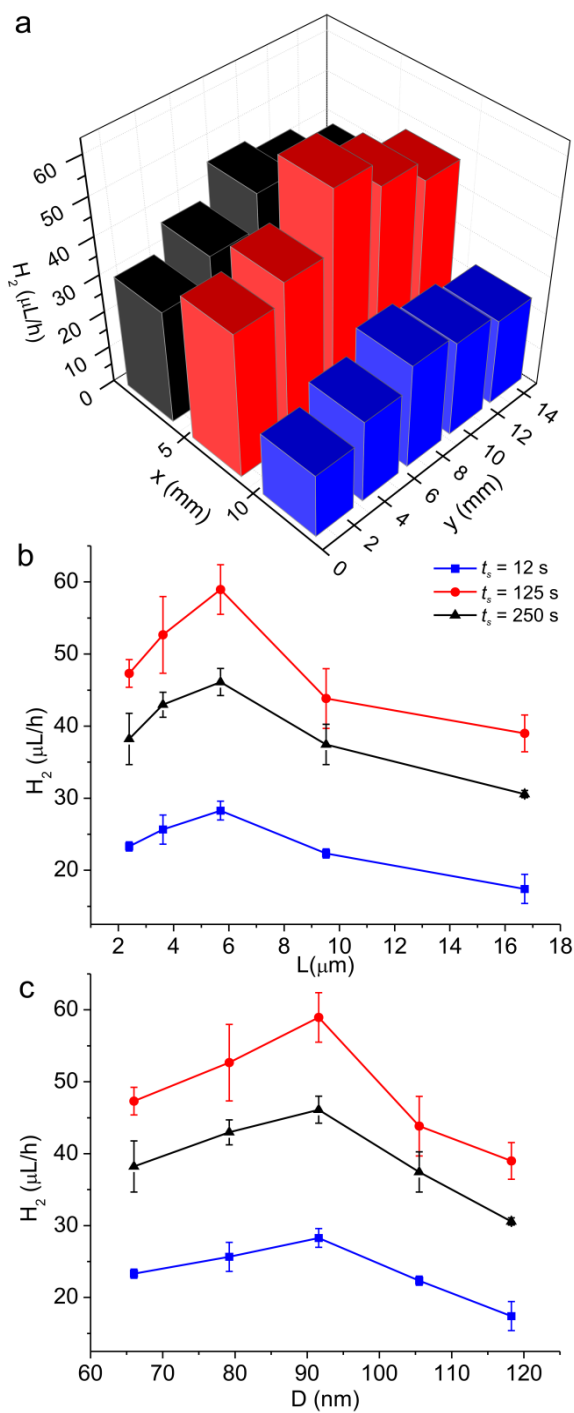

Figure 7. Hydrogen mapping. a) Hydrogen production rate as a function of the position x and y on the gradient. b) Hydrogen pro-duction rate as a function of the length of the tubes L for three Pt loadings. c) Hydrogen production rate as a function of the diameter of the tubes D for three Pt loadings. In all these graphs the blue, red and black plots were obtained for ts = 12, 125 and 250 s, respec-tively.

Conclusion

We demonstrated the straightforward identification of optimal $TiO_2$ NT characteristics for the photocatalytic production of $H_2$ using two-dimensional NT gradient sur-faces. First we showed the fabrication of open $TiO_2$ NT arrays with a broad gradient in length and diameter using bipolar anodization. These arrays were then decorated toposelectively with different amounts of Pt NPs, which yielded catalytically-active layers with a tube gradient in one direction and a gradient of Pt NPs in the other direc-tion. A setup was developed that allows to locally illumi-nate the surface with a UV laser with a precise control over the x- and y-position of the laser spot. The performance of this setup was tested based on kinetic measurements, and used for the mapping of the H2 production at the TiO2 NT gradient surfaces. The surface was screened in x- and y-direction, and the x, y-position of optimal H2 production was spatially identified and related to the TiO2 NT geometry and Pt amount. This yields the hydrogen production rate as a function of the length, the diameter and the amount of Pt. The best tubes had a length of 6 µm, a di-ameter of 92 nm and were decorated with 2.1 wt% (EDX) of Pt NPs.

This low-cost and time-effective technique is a very at-tractive alternative to the use of classical anodization techniques for the optimization of TiO2-based photocata-lytic surfaces. It promises application for various other industrially important reactions such as e.g. CO2 reduction, water depollution and selective catalysis.53-59 This could save a considerable amount of time for the optimization of new photocatalytic systems. Furthermore, this work opens the path for high resolution hydrogen mapping – combin-ing the present concept with advanced optics, mapping with a resolution in the submicrometer range becomes tangible and the photocatalytic rate of a few hundreds TiO2 NTs could be coded by a pixel.

ASSOCIATED CONTENT

Supporting Information

SEM pictures of the top of the layers and the study of the penetra-tion depth of the Pt NPs are provided in supporting information. This material is available free of charge via the Internet at http://pubs.acs.org

AUTHOR INFORMATION

Corresponding Author

*schmuki@ww.uni-erlangen.de

Author Contributions

The manuscript was written through contributions of all authors. All authors have given approval to the final version of the manu-script.

ACKNOWLEDGMENT

This work was supported by a post-doctoral research grant from the Alexander von Humboldt Foundation. ERC and DFG are acknowledged for support. N. T. Nguyen and Dr. Anca Mazare are acknowledged for valuable discussions.

Supplementary information

Low magnification SEM pictures of the top of the tubes

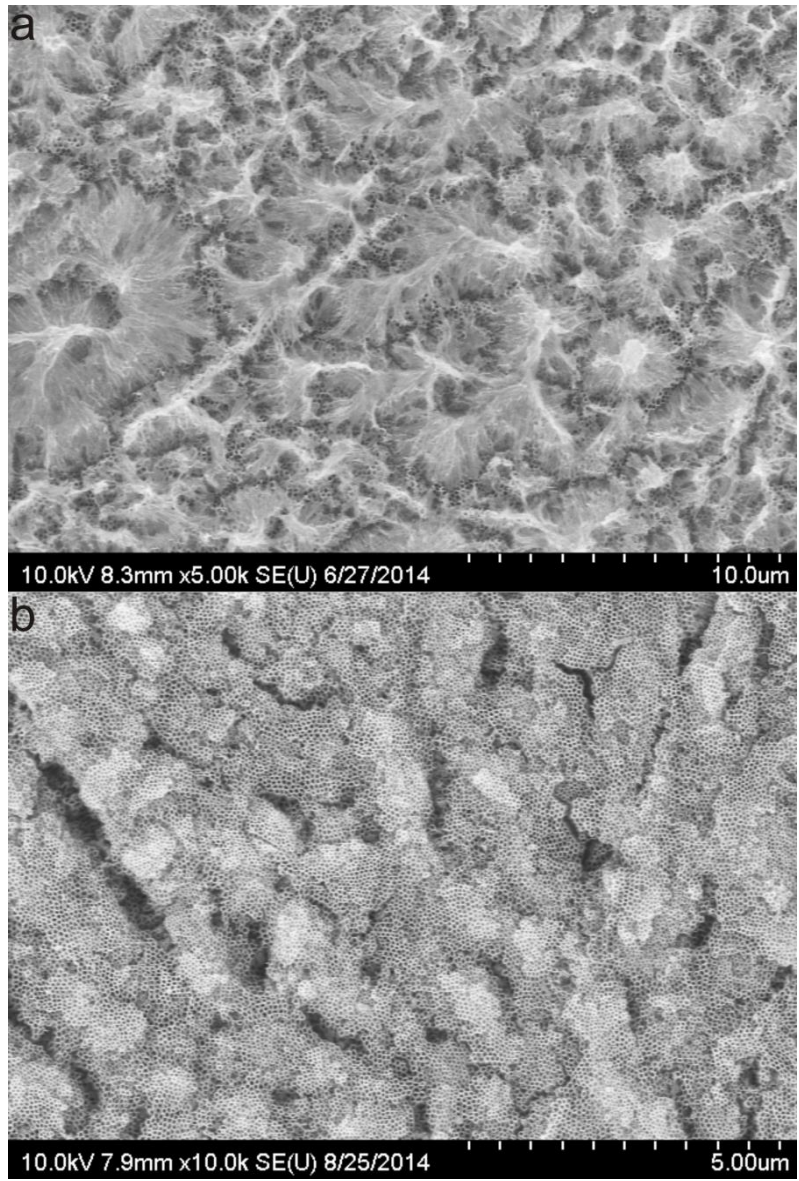

Figure S1. SEM pictures showing typical top views of the TiO$_2$ NT gradient layers before (a) and after (b) opening of the tubes.

## Penetration depth of the Pt NPs inside the layers

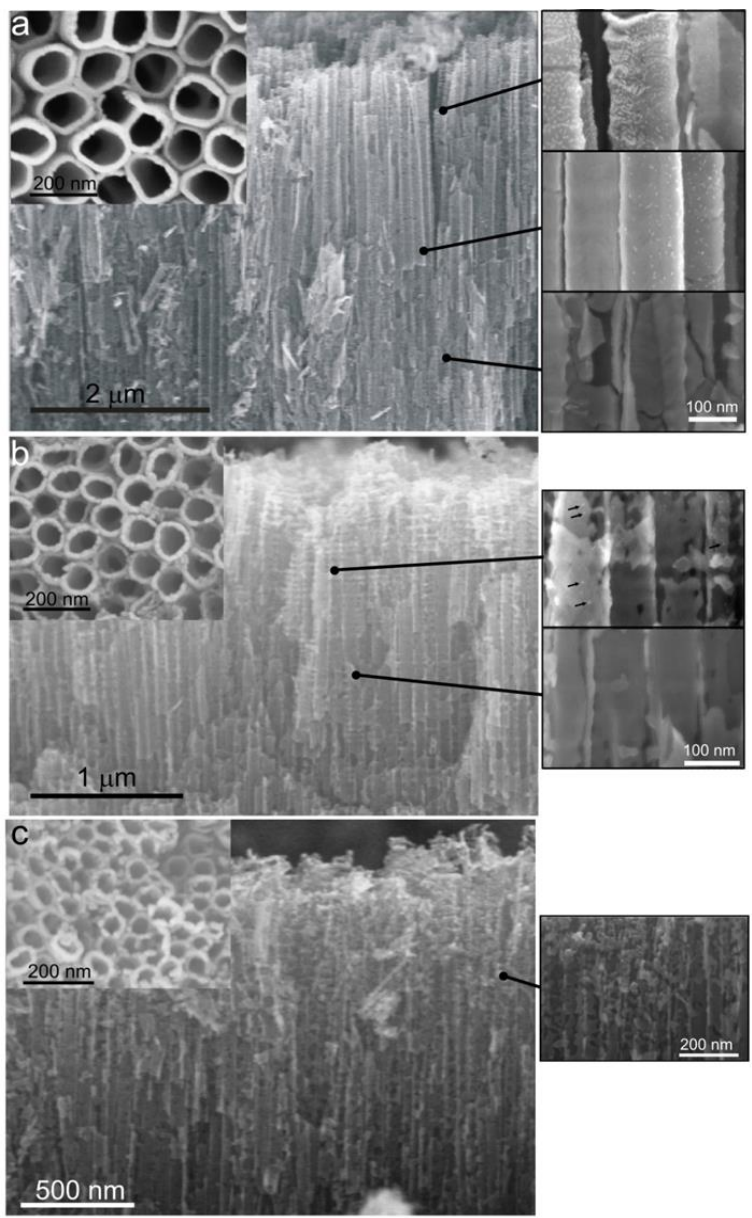

Figure S2. SEM study of the penetration depth of the Pt NPs as a function of the diameter of the tubes, for tubes having diameters of 114 nm (a), 78 nm (b) and 42 nm (c).

Table S1. Penetration depth of the sputtered Pt $P_{Pt}$ as a function of the tube diameter D.

| Position | D (nm) | $P_{Pt}$ (µm) |
|---|---|---|
| a | 114 ±16 | $2.1 < P_{Pt} < 3.2$ |
| b | 78 ±12 | $0.5 < P_{Pt} < 1.2$ |
| c | 42 ±10 | ≈0.5 |

Gas chromatography analysis of H$_2$ obtained for kinetic measurements

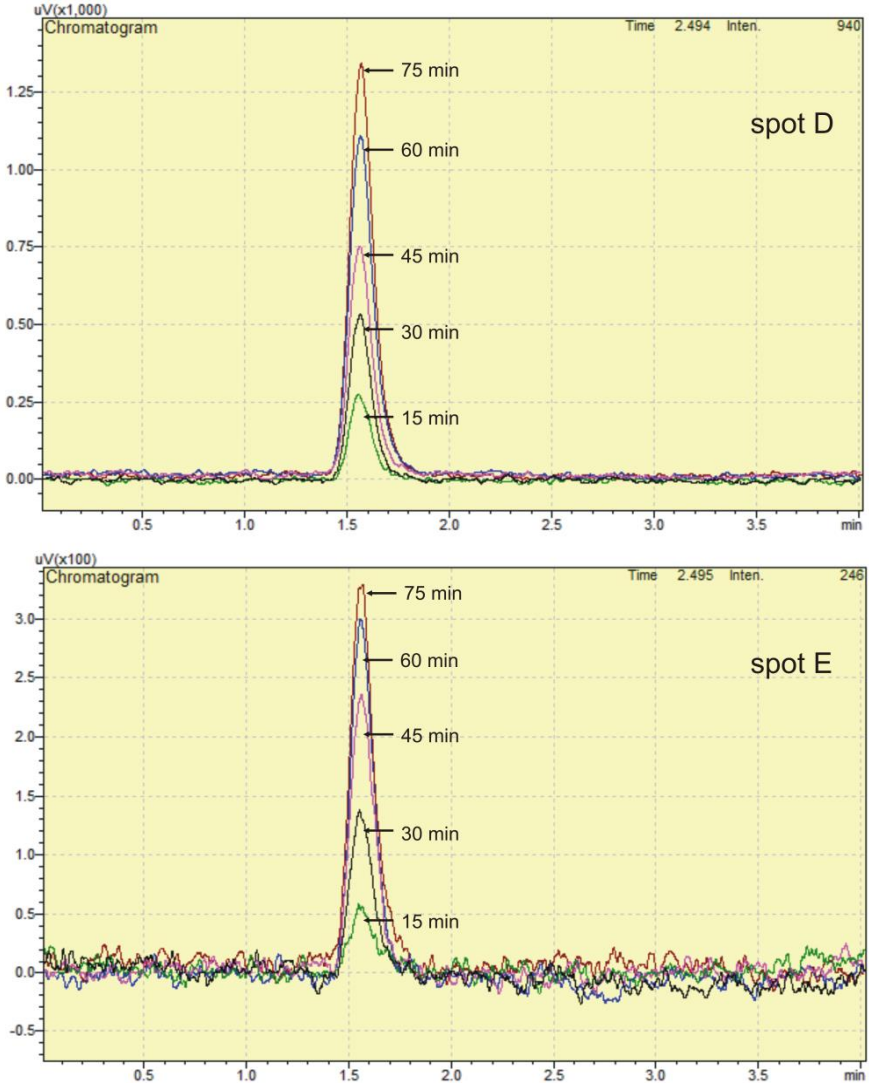

Figure S3. H$_2$ peaks obtained by gas chromatography for the kinetic measurements of spot D (top) and spot E (bottom).